\documentstyle[12pt]{article}

\large
\newcommand{\be}{\begin{eqnarray}}
\newcommand{\ee}{\end{eqnarray}}
\newcommand\del{\partial}

\begin{document}
\setlength{\baselineskip}{21pt}
\pagestyle{empty}
\vfill
\eject
\begin{flushright}
SUNY-NTG-95/50 \\
\end{flushright}

\vskip 1.0cm
\begin{center}
{\bf Scalar susceptibility in QCD and the multiflavor Schwinger model}
\end{center}
\vskip 1.0 cm
\centerline{A. Smilga }
\vskip .2cm
\centerline{ITEP, B. Cheremushkinskaya 25, Moscow 117259, Russia}
\vskip 0.5 cm
\centerline{and}
\vskip 0.5cm
\centerline{J.J.M. Verbaarschot}
\vskip 0.2cm
\centerline{Department of Physics}
\centerline{SUNY, Stony Brook, New York 11794}
\vskip 1cm
\centerline{\bf Abstract}

We evaluate the leading infrared behavior of the scalar susceptibility in QCD
and in the multiflavor Schwinger model for small non-zero quark mass $m$ and/or
small nonzero temperature as well as the scalar susceptibility for the
finite volume QCD partition function.
In QCD, it is determined by one-loop chiral
perturbation theory, with the result that the leading infrared singularity
behaves as $\sim \log m$ at zero temperature and as $\sim T/\sqrt m$ at finite
temperature.
In the Schwinger model with several flavors we use exact results for the
scalar correlation function. We find that the Schwinger model has a phase
transition at $T=0$ with critical exponents that satisfy the standard
scaling relations. The singular behavior of this model depends on the
number of flavors with a scalar susceptibility that
behaves as $\sim m^{-2/(N_f+1)}$.
At finite volume $V$ we show that the scalar susceptibility is proportional
to $1/m^2V$. Recent lattice calculations of this quantity by Karsch
and Laermann are discussed.

\vfill
\noindent
\begin{flushleft}
November 1995
\end{flushleft}
\eject
\pagestyle{plain}

\vskip 1.5cm
\renewcommand{\theequation}{1.\arabic{equation}}
\setcounter{equation}{0}
\noindent

\section{Introduction.}
\vskip 0.5cm
The scalar susceptibility in QCD is defined as
\be
\chi = \int d^4x \langle \sum_{i=1}^{N_f}\bar q_i q_i (x) \ \sum_{i=1}^{N_f}
\bar q_i q_i(0)
\rangle - V \langle \sum_{i=1}^{N_f}\bar q_i
q_i \rangle ^2 = \frac 1V \left . \del^2_m \log Z  \right |_{m= 0},
\label{def}
\ee
where $V$ is the four dimensional Euclidean volume and the averaging is
performed either over the vacuum state or over the thermal ensemble. (In the
Euclidean
approach the latter corresponds to an asymmetric box with imaginary time
extension of  $\beta = 1/T \ll L$, i.e. $V= L^3 \beta$). The definition
(\ref{def}) is for a diagonal mass matrix with equal quark masses.
It is especially interesting to study
this quantity in the neighborhood of the thermal
phase transition point.
It is expected that for QCD with two massless flavors  a
second order phase transition occurs leading to restoration of chiral
symmetry \cite{PW} (see \cite{Varenna,DeTar} for recent reviews) with a
diverging susceptibility. This can be understood simply in terms of
Landau  mean field theory. For a system with order parameter $\eta$ coupled
to external field $h$ with a second order phase transition at $T=T_c$, the
fluctuations of the order parameter are described by the effective potential
\be
V^{\rm eff}(\eta) = A(T-T_c) \eta^2 + B\eta^4 + C\eta h \label {Veff}.
\ee
In QCD $\eta$ is the chiral condensate and $h$ is the quark mass.
At $T=T_c$, the minimum of the potential occurs at $\eta \sim h^{1/3}$
which gives the law $\chi = \del \eta /\del h \sim h^{-2/3}$ for the
susceptibility. On the other hand if $T$ is close to $T_c$ but $T\ne T_c$ and
the external field is weak enough
\be
h \ll |T-T_c|^{3/2}, \label{ht}
\ee
the quartic term in (\ref{Veff})
is irrelevant and the scaling law is $\chi \sim
|T-T_c|^{-1}$ both above and below $T_c$ (the proportionality constant in these
two regions differs by a factor 2).

Recently, on the basis of lattice simulations of the 3-dimensional
Gross-Neveu model it has been suggested that a second order phase transition
involving soft modes consisting out of fermions has critical exponents
given by mean field theory \cite{Kocic-Kogut}. In particular, if this is also
true for QCD with two massless flavors, we get
\be
\langle \bar q q \rangle &\sim& m^{\frac 13}, \label{m13}\\
\chi &\sim& m^{-\frac 23}, \label{m23}
\ee
at $T=T_c$, and
\be
\chi \sim \frac 1{|T-T_c|}
\label{TTc}
\ee
at
\be
\Lambda^{\frac 13} m^{\frac 23} \ll |T-T_c| \ll T_c \label{rangle},
\ee
where $\Lambda$ is a typical hadronic mass scale.
These scaling laws have been reproduced by a simple stochastic matrix model
\cite{JV,Wettig}.
The scalar susceptibility was recently measured for lattice QCD
with two light flavors \cite{Karsch}. They found a diverging susceptibility
at $T=T_c$ with critical exponent $\delta^{-1} = 0.24 \pm 0.03$ which does not
agree with the prediction $\delta^{-1} = 1/3$ of the Landau mean field theory.
Obviously, further numerical
measurements of the critical indices in QCD are highly desirable.

In this paper we will study the scalar susceptibility in three different cases.
First, in section 2 the scalar susceptibility is evaluated for the
multi-flavor massive Schwinger model which  shares many common qualitative
features with $QCD$ and, in particular, it shows a phase transition
with ``restoration'' of chiral symmetry at $T=0$.
All other critical exponents that can be defined in the SM will be evaluated
as well, and it will be shown that
they satisfy the scaling relations modified for a phase transition at zero
temperature. Second, we evaluate the scalar susceptibility in QCD
at low temperatures
using chiral perturbation theory. Third, in order to estimate finite size
effects in lattice calculations, we calculate the scalar susceptibility
in volumes with spatial length below the Compton wave length of the pion.

\setlength{\unitlength}{1mm}
\begin{center}
\begin{picture}(140,30)(10,10)
\thinlines
\put(40,30){\circle{14}}
\put(31.6,28.6){$\times$}
\put(45.4,28.6){$\times$}
\put(25.0,28.6){$\bar q q$}
\put(50.0,28.6){$\bar q q$}
\put(118.0,28.6){$\bar d d$}
\put(82.0,28.6){$\bar u u$}
\put(95,30){\circle{10}}
\put(110,30){\circle{10}}
\put(88.7,28.6){$\times$}
\put(113.3,28.6){$\times$}
\put(40,15){\rm a}
\put(102.5,15){\rm b}
\end{picture}
\end{center}
\setlength{\baselineskip}{15pt}
\noindent
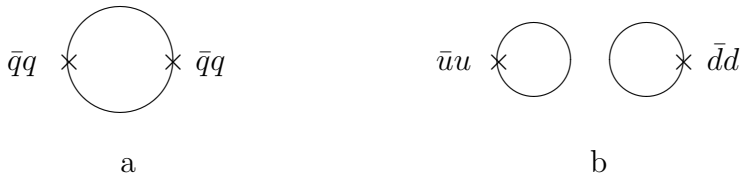
 Fig. 1. Connected (a) and disconnected (b) graphs contributing to the
scalar susceptibility of quarks propagating in a background gluon field.
\setlength{\baselineskip}{21pt}

Before proceeding further, we should note that the scalar susceptibility as
defined in (\ref{def}) involves a quadratic ultraviolet divergence due to
a trivial perturbative graph depicted in Fig. 1a. The situation is the same as
with the chiral condensate which involves a trivial divergent perturbative
contribution $\sim m\Lambda_{UV}^2$. The predictions (\ref{m13},\ref{m23},
\ref{TTc}) hold for the infrared sensitive part of $\langle \bar q q \rangle$
and $\chi$.

The susceptibility (\ref{def}) can be written as the sum of a connected and
a disconnected contribution corresponding to the graphs of Figs. 1a and 1b,
respectively,
\be
\chi = N_f \chi^{\rm con} + N_f^2 \chi^{\rm dis}
\label{chitot}
\ee
The connected contribution to the susceptibility was calculated
in \cite{Stern}. The disconnected contribution is defined by
\be
\chi^{\rm dis} = \int d^4 x \langle \bar u u (x) \bar d d(0) \rangle -
V \langle \bar u u \rangle\langle \bar dd \rangle = \frac 1V
\del_{m_u} \del_{m_d} \log Z,
\label{discon}
\ee
where $m_u$ and $m_d$ are two different quark masses that are put to zero after
differentiation.
We will obtain the latter contribution from the difference of $\chi$ and
$\chi^{\rm con}$.

Both scalar susceptibilities can be expressed in terms of the eigenvalues
$\lambda_k$ of the Dirac operator
\be
\chi^{\rm dis} &=& \frac 1V \left[  \left < \left(\sum_{k} \frac
1{i\lambda_k + m} \right)^2 \right > - \left< \sum_k \frac 1{i\lambda_k + m}
\right> ^2 \right]
,\label{spdis}\\
\chi^{\rm con} &=&
- \frac 1V \left<  \sum_{k} \frac 1{(i\lambda_k+m)^2}
\right>.\label{spcon}
\ee
A related susceptibility is the  pseudo-scalar susceptibility which in
terms of the Dirac eigenvalues is given by
\be
\chi^{\pi} &=& \frac 1V \left<  \sum_{k} \frac 1{\lambda_k^2 +m^2}\right>.
\label{pion-susceptibility}
\ee

\vskip 1.5cm
\renewcommand{\theequation}{2.\arabic{equation}}
\setcounter{equation}{0}
\section{The Schwinger model}
\vskip 0.5 cm
In this section we discuss the Schwinger model (SM)
with $N_f$ light flavors, and
determine the critical exponents from known results for the correlation
functions. For the critical exponents we will follow the convention of
\cite{LL1}.

Critical behavior shows up only in the SM with several flavors. In the standard
Schwinger model ($N_f = 1$), there is no non-anomalous global symmetry to be
broken spontaneously and no reason for the phase transition to occur.
The scalar
susceptibility in the standard SM was determined recently in \cite{Adam}. It
is just a finite constant.

The theory with $N_f > 1$ and zero fermion masses has the global symmetry
$SU_L(N_f) \times SU_R(N_f)$ (much like in $QCD$)
and the potential possibility of its spontaneous breaking with generation of
fermion condensate exists.
Coleman's theorem \cite{Col1} prevents, however, spontaneous breaking of a
continuous
symmetry in 2 dimensions. So, a QCD--like phase transition cannot occur in a 2D
theory at a nonzero temperature.
Nevertheless, it turns out that the dynamics of the
SM at small $T$ is similar to that of a theory with
second order phase transition in the region of temperatures slightly above
critical. One can say that the phase transition {\it does} occur at $T=0$.

The Lagrangian of the model is
\be
{\cal L} = -\frac 14 F_{\mu\nu}^2 + i \sum_{f=1}^{N_f} \bar q_f \gamma_\mu(
\del_\mu - ig A_\mu) q_f - m \sum_{f=1}^{N_f} \bar q_f q_f,
\label{LSM}
\ee
where $g$ is the coupling constant with the dimension of mass, and, for
simplicity, we assumed that all quark masses are equal.

Let us study this
model in the region $m\ll g$. The particle spectrum involves a massive
photon \cite{Weisberger} with mass
\be
\mu_+ \sim g \sqrt{\frac {N_f}{\pi}} + O(m) \label{mup}
\ee
and 'quasi-Goldstone' particles\footnote{It is better to use the term
'quasi-Goldstone' than the term 'pseudo-Goldstone' commonly used for pions
because 'quasi-Goldstone' states in the SM become sterile in the chiral limit
$m\rightarrow 0$. That conforms with Coleman's theorem which forbids the
existence of massless interacting particles in 2 dimensions.} with the
mass \cite{Col2,SM}
\be
\mu_- \sim g^{\frac 1{N_f+1}} m^\frac{N_f}{N_f+1}. \label{mmin}
\ee
This gives us the critical exponent $\mu = N_f/(N_f+1)$.

At nonzero $m$, the chiral $SU_L(N_f) \times SU_R(N_f)$ symmetry
of the massless SM Lagrangian is broken explicitly, and the formation of the
chiral condensate becomes possible. The chiral condensate involves a UV
divergent piece
\be
\langle \bar q q \rangle \sim m \log \Lambda_{UV}, \label{ultr}
\ee
and an infrared contribution (sensitive to the small eigenvalues of the
Euclidean Dirac operator). The latter has been determined in \cite{SM,Hosotani}
with the result
\be
\langle \bar q q \rangle \sim m^{\frac{N_f-1}{N_f+1}} g^{\frac 2{N_f+1}},
\label{condN}
\ee
providing us the critical exponent $\delta = (N_f+1)/(N_f -1)$. The
susceptibility is
  \be
  \chi = \frac{\partial \langle \bar q q \rangle_m}{\partial m} \sim
\left( \frac gm \right)^{2/(N_f+1)}
  \ee
In the region $\mu_+^{-1} \ll |x| \ll \mu_-^{-1}$ the vacuum scalar correlator
is given by \cite{Weisberger,Shrock,SM}
\be
\langle \bar qq(x) \bar qq(0) \rangle_0 \sim \frac {g^{2/N_f}}{|x|^{2-2/N_f}}.
\label{corrN}
\ee
The associated critical exponent is $\zeta = 2 -2/N_f$  (To make contact with
the standard theory of phase transitions, $x$ should be assumed space-like,
but the behavior (\ref{condN}) holds of course for any $x$ due to
Lorentz-invariance ).
At $|x|\gg \mu_-^{-1}$ the correlator levels off at the value of
the square of the chiral condensate (\ref{condN}).

In the region $      \mu_- \ll  T\ll \mu_+$ (weak field limit)
the condensate is given by \cite{Hosotani}
\be
\langle \bar q q \rangle \sim m \left (\frac{g}{T}\right )^{2/N_f},
\label{condint}
\ee
which leads to the susceptibility
\be
\chi \sim \left ( \frac gT\right)^{2/N_f},
\ee
and the critical exponent $\gamma = 2/N_f$.

 The correlation length in this case is just inverse fermion Matsubara
frequency $\sim 1/T$ which gives the critical exponent $\nu = 1$.
For $\mu_- \ll T$ the energy density is that of a gas of massless particles.
Therefore, $\epsilon \sim \frac {\pi }3T^2 $ and the specific heat
for zero field is given by
\be
C = \frac {d\epsilon}{d T} \sim T \label{specheat}
\ee
leading to $\alpha = -1$.

The above results for the critical exponent have been summarized in Table 1.
We also show the critical exponents for mean field theory. Note that for
$N_f = 2$ {\it some} (namely, $\delta$ and $\gamma$), but not all
of the exponents  coincide
with the predictions of MFT.

\begin{table}
\begin{center}
\begin{tabular}{c c c}
exponent  & MFT          &SM \\ \hline
$\alpha$  &   0          &  $-1$ \\
$\beta$   &   $\frac 12$ & --- \\
$\gamma$  &   1          &   $\frac 2{N_f}$ \\
$\delta $ &  3           &  $\frac{N_f+1}{N_f-1}$ \\
$\varepsilon$&  0           &  --- \\
$\mu$     &  $\frac 13$  & $\frac{N_f}{N_f+1}$\\
$\nu$     & $\frac 12 $  & 1        \\
$\zeta$   &   0          &$ 2 - \frac 2{N_f} $\\
\end{tabular}
\end{center}
\caption{ The critical exponents
for mean field theory (MFT) and
the Schwinger model (SM). Conventions are as in Landau and Lifshitz [5].}
\end{table}

It is instructive to write down an effective non-local lagrangian from where
these values of critical exponents can be inferred:
  \be
  \label{Leta}
{\cal L} = A\eta (\Delta + BT^2)^{1/N_f} \eta + C \eta^{\frac{2N_f}{N_f -1}}
+ D\eta h
  \ee
Of course, (\ref{Leta}) is just a shorthand for the values of critical indices
obtained and should be understood as such.

In standard theory of second order phase transitions with a nonzero critical
temperature the above 8 critical exponents satisfy 5 universal thermodynamic
relations and the hyper-scaling relation.
The fact that the critical temperature is zero  brings about a
number of modifications compared to the standard theory of phase transitions:

\begin{itemize}
\item The critical exponent $\beta$ refers to the broken phase and therefore
      cannot be defined.
\item In the strong field limit $h \gg t^{\nu/\mu}$ the partition function
      depends on the temperature as
      $\propto \exp\{- \frac{h^{\mu/\nu}}{t}\}$. Then the critical exponent
      $\varepsilon$ becomes singular for $t\rightarrow 0$ and
      cannot be defined.
      Specifically, for the SM the
      free energy of the gas of bosons with the mass of order $\mu_-$
      remains exponentially  small until
      $t \sim \mu_- \propto m^\mu$ ($\nu = 1$ in our case).
\item Because $c_p = -T \del^2\Phi/\del T^2$ and $T \equiv T - T_c$ (with
$T_c=0$), an extra
power of $T$  emerges leading to the Rushbrooke scaling relation
\be \alpha+ 2\beta + \gamma = 1
\label{rush} \ee
instead of 2. After elimination of $\beta$ and $\epsilon$ we have only 3
instead of the usual 5 relations:
\be
   \label{alwrong}
\alpha + \frac{\gamma(\delta + 1)}{\delta - 1}  &=& 1\\
\nu(2-\zeta) &=& \gamma. \label{s4}\\
\mu(1+\gamma-\alpha)&=& 2\nu \label{ss4}
\ee

\item The hyperscaling relation follows from the condition that the total
      free energy is of the order of $T V/\xi^d$, where $V$ is the spatial
      volume of dimension $d$ and $\xi$ is the correlation length. For
      $T_c = 0$ we obtain
  \be \nu d = -\alpha \label{hyper}\ee
      instead of $2-\alpha$.
\end{itemize}
The meaning of the hyperscaling relation is that loop corrections to the
Green's
functions estimated from the effective lagrangian (\ref{Leta}) are of the same
order, as far as powers are concerned, as the tree expressions. We emphasize
again that one cannot perform {\it serious} loop calculations with the
effective lagrangian (\ref{Leta}). If trying to do so, logarithmic factors
would
appear which would shift ``tree level'' values of exponents.
\renewcommand{\theequation}{3.\arabic{equation}}
\setcounter{equation}{0}
\vskip 1.5cm
\section{QCD at low temperatures}
\vskip 0.5cm

The primary interest of the quantity (\ref{def}) is that its critical behavior
can provide information on the physics of the phase transition in QCD. However,
our second remark is that the leading infrared behavior of $\chi$ can be
determined $exactly$ in the low temperature region. The proper technique to
extract it is chiral perturbation theory \cite{CPT}.
\vskip -0.5cm
\begin{picture}(150,30)(0,10)
\put(76.24,20.21){.}
\put(75.77,21.98){.}
\put(74.85,23.56){.}
\put(73.56,24.85){.}
\put(71.98,25.77){.}
\put(70.21,26.24){.}
\put(68.39,26.24){.}
\put(66.62,25.77){.}
\put(65.04,24.85){.}
\put(63.75,23.56){.}
\put(62.83,21.98){.}
\put(62.36,20.21){.}
\put(62.36,18.39){.}
\put(62.83,16.62){.}
\put(63.75,15.04){.}
\put(65.04,13.75){.}
\put(66.62,12.83){.}
\put(68.39,12.36){.}
\put(70.21,12.36){.}
\put(71.98,12.83){.}
\put(73.56,13.75){.}
\put(74.85,15.04){.}
\put(75.77,16.62){.}
\put(76.24,18.39){.}
\put(61.4,18.55){$\times$}
\put(75.3,18.55){$\times$}
\end{picture}

\noindent
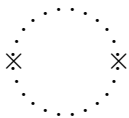
Fig. 2. Pseudo-Goldstone loop determining the scalar susceptibility.

 The leading infrared behaviour is
determined by the graph in Fig. 2 involving a loop of quasi-massless
pseudo-Goldstone bosons. The effective low-energy Lagrangian of QCD has the
form
\be
{\cal L}^{\rm eff} = \frac 14 F_\pi^2 {\rm Tr } (\del_\mu U^\dagger)(\del_\mu
U)+\Sigma{\rm Re}{\rm Tr}\{{\cal M} U\} + \cdots, \label{Lchi}
\ee
where $U=\exp\{ 2i\phi^a t^a/F_\pi\}$ and $\phi^a$ are the pseudo-Goldstone
fields. The chiral condensate is denoted by $\Sigma = |\langle \bar q
q\rangle_0 |$ (no summing over colors assumed), and ${\cal M}$
is the quark mass
matrix. The partition function with ${\cal M}={\rm diag}(m,m,\cdots,m)$
has been calculated by Gasser and Leutwyler \cite{Gasser}. Their expression for
the free energy density of lukewarm pion gas is

\begin{equation}
f = \epsilon_0(M_\pi) - \frac {N_f^2-1}2 g_0(T, M_\pi)
+ \frac {N_f^2-1}{4N_f}\ \frac {M_\pi^2}{F_\pi^2} g_1^2(T, M_\pi)
\label{FGL}
\end{equation}
where
  \be
  \label{g0}
  g_0(T, M_\pi) = - \frac T{\pi^2} \int p^2 dp \ln  [1 - \exp (- E/T )],
  \ee
 $E = \sqrt{p^2 + M_\pi^2}$, and
  \be
  \label{g1}
g_1(T, M_\pi) \ =\ \frac 1{2\pi^2} \int \frac{p^2 dp}{E[\exp(E/T)-1]} =
-\frac{\partial g_0(T, M_\pi)}{\partial M_\pi^2}
  \ee
Here, $M_\pi$ stands for the common mass of the $N_f^2 -1 $ pseudo-Goldstone
modes given by the Gellmann-Oakes-Renner relation $F_\pi^2 M_\pi^2 = 2m\Sigma$.

Substituting in Eq.(\ref{FGL}) the expansion \cite{Weldon,Gerber}
\be
g_0 = \frac{\pi^2}{45} T^4 -
\frac{T^2 M_\pi^2}{12} + \frac {TM_\pi^3}{6\pi},
\label{expans}
\ee
we obtain
  \be
f &=& - \frac{\pi^2}{90}(N_f^2-1) T^4
-\frac {N_f}2 F^2_\pi M^2_\pi\left (1- \frac
{N_f^2-1}{12N_f}\frac{T^2}{F_\pi^2}
-\frac{N_f^2-1}{288N_f^2}\frac{T^4}{F_\pi^4}\right ) \nonumber\\
&-&\frac{N_f^2-1}{12\pi}T M^3_\pi
\left(1+ \frac 1{8N_f} \frac {T^2}{F_\pi^2} \right )
\nonumber\\
&-&\frac{N_f^2 -1}{64\pi^2} M^4_\pi
\log \left (\frac {\Lambda^2}{M_\pi^2 }\right),
\label{free}
\ee

Consider first the case of zero temperature. For the infrared singular
contribution to the scalar susceptibility we obtain
\be
\chi^{IR} = \frac{N_f^2-1}{8\pi^2 }\left (\frac {\Sigma}{F_\pi^2} \right )^2
\log \frac{\Lambda^2}{M_\pi^2},
\label{chiQCD}
\ee
which was first obtained in \cite{Shifman} from an analysis of the spectral
function in the scalar channel and the PCAC hypothesis.

The infrared singular
connected contribution to the susceptibility was calculated in \cite{Stern}
(One should just substitute 1 for ${\rm Tr} \{t^at^b\} = \delta^{ab}/2$ in
Eq.(2.13) of Ref.\cite{Stern}.). The result is
\be
\chi^{\rm IR\, con} = \frac{N_f^2-4}{16\pi^2 N_f}\left (\frac {\Sigma}{F_\pi^2}
\right )^2 \log \frac{\Lambda^2}{M_\pi^2}.
\ee
Using (\ref{chitot}) we immediately obtain the disconnected contribution
\be
\chi^{\rm IR\, dis} = \frac{N_f^2 + 2}{16\pi^2 N_f^2}\left (\frac
{\Sigma}{F_\pi^2}
\right )^2 \log \frac{\Lambda^2}{M_\pi^2}.
\ee
Instead of using the partition function of Gasser and Leutwyler we can
calculate the susceptibility also using the same technique as in \cite{Stern}.
Choosing ${\cal M}={\rm diag}(m,m,\cdots,m)$, one immediately obtains the
vertex
\be
\langle 0| \sum_f \bar q_f q_f | \phi^a \phi^b \rangle = \frac
{2\Sigma}{F_\pi^2} \delta^{ab},
\label{vertex}
\ee
which, by evaluation of the diagram in Fig. 2, reproduces the result
(\ref{chiQCD}).

The infrared singular contribution to the susceptibility af finite temperature
can be obtained directly from (\ref{free}) as well
\be
\chi_T^{IR} = \frac{(N_f^2-1)}{ 4\pi } \frac T{\sqrt{2m}}  \left (
\frac {\Sigma}{F_\pi^2}\right )^{3/2},
\label{chiTQCD}
\ee
where the Gellmann-Oakes-Renner relation has been used.
The connected contribution to the susceptibility was not calculated in
\cite{Stern}. However, the  zero temperature result of Ref.\cite{Stern}
can be extended immediately to finite $T$ by making
the substitution
\be
\int \frac {d^4p}{(2\pi)^4} \rightarrow T \sum_n\int \frac{d^3p}{(2\pi)^3},
\label{Matsint}
\ee
where the sum is over all Matsubara frequencies ($p_0 = 2\pi n/\beta$).
The infrared singular part comes only from term $n=0$ and
the result is
\be
\chi_T^{\rm IR\, con} = \frac{N_f^2-4}{ 8\pi N_f} \frac T{\sqrt{2m}}
\left ( \frac {\Sigma}{F_\pi^2}\right )^{3/2}.
\label{chiTQCDcon}
\ee
This implies that at finite temperature the spectrum of the Dirac
operator is nonanalytic  for small eigenvalues.
The disconnected contribution is
\be
\chi_T^{\rm IR\, dis} = \frac{N_f^2+2}{ 8\pi N_f^2} \frac T{\sqrt{2m}}
\left ( \frac {\Sigma}{F_\pi^2}\right )^{3/2}.
\label{chiTQCDdis}
\ee

The relations (\ref{chiTQCD}, \ref{chiTQCDcon}, \ref{chiTQCDdis})
are quite analogous to the
relations for the magnetic susceptibility for ferromagnets known for a long
time. They are also determined by a loop of pseudo-Goldstone particles (the
magnons) depicted in Fig. 2 and have
the same behavior\footnote{Recently, a full
nonlinear effective Lagrangian has been constructed \cite{Leut} in a way which
makes the analogy between the theory of ferromagnets and CPT most transparent.}
\cite{LL2}
\be
\chi^{\rm magnon} = \left . \frac {\del M}{\del H}\right |_{H=0} \sim \frac
T{\sqrt H}. \label{chimagn}
\ee

\begin{center}
\begin{picture}(140,40)(10,10)
\put(40,26.8){\circle*{1}}
\put(31.4,18.55){$\times$}
\put(45.3,18.55){$\times$}
\put(81.4,18.55){$\times$}
\put(95.3,18.55){$\times$}
\put(46.24,20.21){.}
\put(45.77,21.98){.}
\put(44.85,23.56){.}
\put(43.56,24.85){.}
\put(41.98,25.77){.}
\put(40.21,26.24){.}
\put(38.39,26.24){.}
\put(36.62,25.77){.}
\put(35.04,24.85){.}
\put(33.75,23.56){.}
\put(32.83,21.98){.}
\put(32.36,20.21){.}
\put(32.36,18.39){.}
\put(32.83,16.62){.}
\put(33.75,15.04){.}
\put(35.04,13.75){.}
\put(36.62,12.83){.}
\put(38.39,12.36){.}
\put(40.21,12.36){.}
\put(41.98,12.83){.}
\put(43.56,13.75){.}
\put(44.85,15.04){.}
\put(45.77,16.62){.}
\put(46.24,18.39){.}
\put(43.21,31.43){.}
\put(42.54,32.95){.}
\put(41.30,34.06){.}
\put(39.72,34.58){.}
\put(38.06,34.40){.}
\put(36.62,33.57){.}
\put(35.65,32.23){.}
\put(35.30,30.60){.}
\put(35.65,28.97){.}
\put(36.62,27.63){.}
\put(38.06,26.80){.}
\put(39.72,26.62){.}
\put(41.30,27.14){.}
\put(42.54,28.25){.}
\put(43.21,29.77){.}
\put(96.24,20.21){.}
\put(95.77,21.98){.}
\put(94.85,23.56){.}
\put(93.56,24.85){.}
\put(91.98,25.77){.}
\put(90.21,26.24){.}
\put(88.39,26.24){.}
\put(86.62,25.77){.}
\put(85.04,24.85){.}
\put(83.75,23.56){.}
\put(82.83,21.98){.}
\put(82.36,20.21){.}
\put(82.36,18.39){.}
\put(82.83,16.62){.}
\put(83.75,15.04){.}
\put(85.04,13.75){.}
\put(86.62,12.83){.}
\put(88.39,12.36){.}
\put(90.21,12.36){.}
\put(91.98,12.83){.}
\put(93.56,13.75){.}
\put(94.85,15.04){.}
\put(95.77,16.62){.}
\put(96.24,18.39){.}
\put(81.91,20.13){.}
\put(81.24,21.65){.}
\put(80.00,22.76){.}
\put(78.42,23.28){.}
\put(76.76,23.10){.}
\put(75.32,22.27){.}
\put(74.35,20.93){.}
\put(74.00,19.30){.}
\put(74.35,17.67){.}
\put(75.32,16.33){.}
\put(76.76,15.50){.}
\put(78.42,15.32){.}
\put(80.00,15.84){.}
\put(81.24,16.95){.}
\put(81.91,18.47){.}
\put(40,5){\rm a}
\put(88.5,5){\rm b}
\end{picture}
\end{center}
\setlength{\baselineskip}{15pt}
\noindent
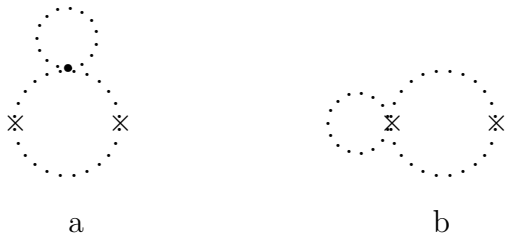
Fig. 3. Two-loop contributions to the scalar susceptibility: a) mass
renormalization; b) vertex renormalization.
\setlength{\baselineskip}{21pt}

The relations  (\ref{chiTQCD}, \ref{chiTQCDcon}, \ref{chiTQCDdis})
hold in the low-temperature region
\be
M_\pi \ll T \ll T_c. \label{range1}
\ee
This makes a direct comparison with recent lattice calculations \cite{Karsch}
impossible.
At not so low temperatures, besides the graph in Fig. 2, also higher order
graphs in chiral perturbation theory contribute. The relevant two-loop graphs
are depicted in Fig. 3, but the results for the susceptibility can be obtained
directly from the partition function as well (\ref{free}),
\be
\chi_T^{IR} = \frac{(N_f^2-1)}{ 4\pi} \frac T{\sqrt{2m}} \left (
\frac {\Sigma}{F_\pi^2}\right )^{3/2} \left( 1 +\frac 1{8N_f} \frac
{T^2}{F_\pi^2} \right).
\label{chiTQCD2lp}
\ee
The two loop temperature dependence can be absorbed in the one-loop
temperature modification of the
condensate and the pion decay constant
as obtained in \cite{Gasser}
\be
\Sigma(T) &=& \Sigma(0) \left(1-\frac {N^2_f-1}{12 N_f} \frac {T^2}{F_\pi^2}
\right),\\
F_\pi(T) &=& F_\pi(0) \left(1-\frac {N_f}{24} \frac {T^2}{F_\pi^2}\right ).
\ee

It is not clear what happens to next order in $T$. Temperature corrections to
the condensate have been found to three-loop level in \cite{Gerber}. Two-loop
effects in $M_\pi^2(T)$ have been extracted in \cite{Goity,Schenk}, but
$F_\pi^2(T)$ at the two-loop level is currently unknown. It is not clear
whether one can just substitute the temperature dependent $\Sigma$ and
$F_\pi^2$ in (\ref{chiTQCD}) to all orders in the temperature expansion.
The same question can be asked concerning the Gellmann-Oakes-Renner relation.
Taking into account corrections of order $T^2/F_\pi^2$, it also holds
at nonzero temperatures. Whether it holds at higher orders is an open
question \cite{Leut1}.

\vskip 1.5cm
\renewcommand{\theequation}{4.\arabic{equation}}
\setcounter{equation}{0}
\section{The scalar susceptibility in finite volumes}
\vskip 0.5cm
The results (\ref{chiQCD}, \ref{chiTQCD}, \ref{chiTQCDcon}, \ref{chiTQCDdis})
are valid in the thermodynamic limit which means that the spatial length of
the box where the theory is defined is much larger than the pion Compton
wave length. However, realistic boxes used in lattice calculations can
never be made so large if the pion mass is small enough (which is in turn
necessary for the quantitative analytic predictions to be possible).
In this section we consider the opposite limit
\be
\Lambda^{-1} \ll L \ll M_\pi^{-1} \sim \frac 1{\sqrt{m\Lambda}}
\label{range2}
\ee
where the susceptibility can be found from the exact results for the
finite-volume partition function of \cite{LS}.
In this range
the susceptibility is expected to be determined by finite volume effects and
be of order
\be
\frac{ \Sigma}m\left (a_0 + a_1 \frac 1{mV\Sigma} + O\left(\frac
1{(mV\Sigma)^2}
\right) \right )
\label{rangesus}
\ee
which is much larger than its thermodynamic limit
$\sim \Lambda^2$. If $a_0 \ne 0$, the result becomes
independent of the volume suggesting that it holds for $V\rightarrow \infty$
outside of the range (\ref{range2}) provided that we are sufficiently close
to the chiral limit so that $\Sigma/m \gg \Lambda^2$. This
is indeed what happens
for the pion susceptibility but not for the scalar susceptibility (see below).
In general, the results of this section have the
same status as in Ref.\cite{LS}. They say little about the properties of
the theory in the physical infinite volume limit but can be used to test
the validity of numerical calculations in $QCD$ which are performed for finite
volumes.

The theoretically simplest susceptibility is the pseudoscalar (or
pion) susceptibility. Using the Banks-Casher relation \cite{Banks-Casher},
it can be related immediately to the chiral condensate,
(see eq. (\ref{pion-susceptibility}))
\be
\chi^{\pi} = \frac \Sigma m,
\ee
in the limit that $\Sigma m V \gg 1$. The validity of this result extends
into the thermodynamic limit outside of the range (\ref{range2}).

As follows immediately from (\ref{spcon},\ref{spdis})
$\chi^{\rm con}$ can be calculated from
average spectral density
  \be
  \label{rho}
  \rho(\lambda) = \langle \omega(\lambda, A) \rangle = \langle
\frac 1V \sum_n \delta(\lambda - \lambda_n) \rangle
\label{onepoint}
  \ee
 On the other hand,
$\chi^{\rm dis}$ can be expressed into
the connected two-point level correlation function
\be
\rho_c(\lambda,\lambda') = V \left[ \langle \omega(\lambda)\omega(\lambda')
\rangle
-\rho(\lambda)\rho(\lambda') \right].
\label{twopoint}
\ee
If, apart from the pairing $\pm \lambda_k$,
the eigenvalues of the Dirac operator are uncorrelated \cite{NVZ},
i.e. if $<f(\lambda_n) g(\lambda_m)> = <f(\lambda_n)> <g(\lambda_m)>$
 for $n \neq m$ and $\lambda_n, \,\lambda_m > 0$, we
have in the limit of large $V$ and for positive $\lambda, \lambda' $
\be
\rho_c(\lambda,\lambda') =
\rho(\lambda) \delta(\lambda-\lambda').
\label{uncorf}
\ee
Using the $U_A(1)$ symmetry of the Dirac spectrum the disconnected
susceptibility (\ref{spdis}) can
be written as
\be
\chi^{\rm dis} =  \int_0^\infty d\lambda \int_0^\infty d\lambda' \frac {4 m^2
\rho_c(\lambda,\lambda') }{(\lambda^2 +m^2)(\lambda'^2 + m^2)}
\ee
which after insertion of the correlation function (\ref{uncorf}) leads to
\be
\chi^{\rm IR\, dis} = \frac {\pi \rho(0)}m = \frac {\Sigma}{m},
\label{uncor}
\ee
where the second equality is the Banks-Casher relation \cite{Banks-Casher}.

Because the diagonalization of the Dirac operator induces correlations
between the eigenvalues, we expect  quite a different prediction from the
finite volume partition function.

For $N_f = 2$ the finite volume partition function is known \cite{LS}.
 For $\theta =0$,
\be
Z = \frac 2{V\Sigma(m_u+m_d)} I_1(V\Sigma(m_u+m_d)) \label{Z}
\ee
where $V= L^3/T$ in the 4D Euclidean volume.
The susceptibilities $\chi^{\rm con}$ and $\chi^{\rm dis}$ can be
disentangled by differentiating with respect to {\it different} quark masses.
Because the partition function depends on the quark masses via the sum
$m_u+m_d$ we find
\be
\chi^{\rm con} = 0.
\label{chiVcon}
\ee
In the limit $\kappa \equiv V\Sigma m \gg 1$ the disconnected contribution
simplifies
\be
\chi^{\rm dis} = \frac 3{8m^2 V}.
\label{chiVdis}
\ee

In present day lattices the zero mode states are mixed with the
much larger number of nonzero mode states. Therefore, the
partition function is effectively calculated in the sector $\nu =0$. Existing
numerical calculations in the instanton liquid model
\cite{inst} are also done for $\nu = 0$.
In a sector with fixed topological charge
the finite volume partition function has been calculated analytically for
an arbitrary number of flavors with equal mass \cite{LS}. To
leading order in $\kappa^{-1}$, the partition function in the sector
of zero topological charge is given by
\be
Z_{\nu =0}^{\rm eff} \sim \frac{\exp(N_f\kappa)}{\kappa^{N_f^2/2}}.
\ee
For $\kappa \gg 1 $ we find
\be
 N_f \chi^{\rm con} + N_f^2 \chi^{\rm dis} =\frac {N_f^2}2 \frac 1{m^2 V}.
\label{chifin}
\ee
In general we have not been able to calculate the connected and disconnected
contributions to the susceptibility separately. However,
for $N_f = 2$ the partition function is known for different quark masses
\cite{LS}. For $\nu = 0$ we find
\be
Z_{\nu = 0} = \frac 1{2\pi} \int_0^{2\pi} d\theta \frac{2I_0(V\Sigma
(m_u^2 +m_d^2 +2m_um_d \cos \theta)^{1/2})}
{V\Sigma (m_u^2 +m_d^2 +2m_um_d \cos \theta)^{1/2}},
\ee
which allows us to calculate the connected and disconnected pieces of
the susceptibility separately. The result for $\kappa \gg 1$ is
$\chi^{\rm con} = 1/2 m^2 V$ and $\chi^{\rm dis} = 1/4m^2 V$. For $N_f = 0$
the two contributions to the susceptibility can be obtained from the spectral
density and the two level spectral correlation function which can be derived
from chiral random matrix theory. The result for $ \kappa \gg 1$
is \cite{JVC}  $\chi^{\rm con} = 0$ and $\chi^{\rm dis} = 1/4m^2 V$.
Our conjecture for arbitrary $N_f$ and $\nu = 0$ consistent with all the above
results is
\be
\chi^{\rm con} &=& \frac{N_f}4 \frac 1{m^2 V}F^{\rm con} (\kappa),\\
\chi^{\rm dis} &=& \frac 1{4m^2 V}F^{\rm dis} (\kappa),
\ee
in agreement with the simplest possible flavor dependence consistent
with (\ref{chifin}). Both $F^{\rm con} (\kappa)$
and $F^{\rm dis} (\kappa)$ approach 1 for $\kappa\gg 1$, but will in general
depend on $N_f$ for finite $\kappa$.
The disconnected
contribution is suppressed by a factor $1/mV\Sigma$ with respect
to the result for uncorrelated eigenvalues. The coefficient $a_0$ in
(\ref{rangesus}) turns out to be zero
which is in agreement with the fact that there
are no massless scalar particles.

The quark mass dependence of the scalar susceptibility
has been calculated by lattice QCD simulations
only for relatively large quark masses \cite{Karsch}.
In this work  with $m^2V \approx 1$ (in units of the lattice spacing)
a quark mass dependence of $\sim 1/m$ is not only found at the critical point
but also at lower temperatures where the susceptibility levels off
at a significantly lower value. This result is in between the finite volume
prediction (\ref{chifin}) and the result from chiral perturbation theory
(\ref{chiTQCD}) with  mass dependence of $1/m^2 $ and $ 1/\sqrt{m}$,
respectively.
The lattice results for $\chi^{\rm dis}$ agree with the mass dependence
(\ref{uncor}) suggesting that the eigenvalues of the
Dirac operator are only weakly correlated. Two different types of
correlations can be considered, namely correlations between
eigenvalues corresponding to different gauge field configurations and
correlations between eigenvalues obtained from the same lattice
gauge configuration. It has been shown \cite{Halasz} that lattice
eigenvalues show strong spectral correlations (the latter type), but this
does not exclude the possibility that correlations between eigenvalues
corresponding to different members of the ensemble are absent. In random
matrix theory it has been shown \cite{Pandey} that spectral averages
and ensemble averages are the same. The exciting possibility that this
type of generalized ergodicity does not hold for the lattice Dirac
eigenvalues deserves further attention.

\vskip 1.5cm
\noindent
{\bf Acknowledgements}
\vglue 0.4cm
 The reported work was partially supported by the US DOE grant
DE-FG-88ER40388 and the INTAS grant 93-0283.
We are much indebted to Y. Hosotani, A. Larkin, V. Lebedev,
and E. Shuryak for
illuminating discussions and to H. Leutwyler for helpful correspondence.
One of us (A.V.S) acknowledges  warm
hospitality extended to him by the Nuclear Theory Group of the
State University of New York at Stony Brook were this work was done.

\vfill
\eject
\newpage
\setlength{\baselineskip}{15pt}

\bibliographystyle{aip}

\vfill
\eject
\end{document}